\documentclass[sigconf]{acmart}

\AtBeginDocument{%
  \providecommand\BibTeX{{%
    \normalfont B\kern-0.5em{\scshape i\kern-0.25em b}\kern-0.8em\TeX}}}

\copyrightyear{Accepted and under publication for 2023}
\acmYear{2023}
\setcopyright{none}
\acmConference[CSCW '23 Companion]{Computer Supported Cooperative Work and Social Computing}{October 14--18, 2023}{Minneapolis, MN, USA}
\acmBooktitle{Computer Supported Cooperative Work and Social Computing (CSCW '23 Companion), October 14--18, 2023, Minneapolis, MN, USA}
\acmPrice{15.00}
\acmDOI{10.1145/3584931.3607017}
\acmISBN{979-8-4007-0129-0/23/10}

\begin{document}

\title{Affective Affordance of Message Balloon Animations: An Early Exploration of AniBalloons}

\author{Pengcheng An}
\orcid{0000-0002-7705-2031}
\author{Chaoyu Zhang}
\affiliation{%
  \institution{School of Design, Southern University of Science and Technology}
  \streetaddress{}
  \city{Shenzhen}
  \country{China}
  \postcode{}
}
\email{anpc@sustech.edu.cn}

\author{Haichen Gao}
\affiliation{%
  \institution{School of Creative Media, City University of Hong Kong}
  \streetaddress{}
  \city{Hong Kong}
  \country{China}}
\email{haichgao-c@my.cityu.edu.hk}

\author{Ziqi Zhou}
\affiliation{%
  \institution{School of Computer Science, University of Waterloo}
  \city{Waterloo}
  \state{Ontario}
  \country{Canada}
}
\email{z229zhou@edu.uwaterloo.ca}

\author{Linghao Du}
\author{Che Yan}
\affiliation{%
 \institution{Huawei Canada}
 \streetaddress{}
 \city{Markham}
 \state{Ontario}
  \country{Canada}}
\email{{che.yan,linghao.du}@huawei.com}

\author{Yage Xiao}
\affiliation{%
  \institution{School of Design, Southern University of Science and Technology}
  \city{Shenzhen}
  \state{}
  \country{China}
}
\email{kayleexiao1109@gmail.com}

\author{Jian Zhao}
\affiliation{%
\institution{School of Computer Science, University of Waterloo}
  \city{Waterloo}
  \state{Ontario}
  \country{Canada}}
\email{jianzhao@uwaterloo.ca}
\authornote{Jian Zhao is the corresponding author.}

\renewcommand{\shortauthors}{An and Zhao, et al.}

\begin{abstract}
  We introduce the preliminary exploration of \textit{AniBalloons}, a novel form of chat balloon animations aimed at enriching nonverbal affective expression in text-based communications. AniBalloons were designed using extracted motion patterns from affective animations and mapped to six commonly communicated emotions. An evaluation study with 40 participants assessed their effectiveness in conveying intended emotions and their perceived emotional properties. The results showed that 80\% of the animations effectively conveyed the intended emotions. AniBalloons covered a broad range of emotional parameters, comparable to frequently used emojis, offering potential for a wide array of affective expression in daily communication. The findings suggest AniBalloons' promise for enhancing emotional expressiveness in text-based communication and provide early insights for future affective design.
\end{abstract}




\maketitle

\section{Introduction}

Text messages have become an essential form of communication in today's world, with more than five billion people using Short Message Services (SMS) globally. The widespread adoption of mobile and wearable devices, as well as messaging apps such as WhatsApp, Facebook Messenger, and WeChat, have further contributed to the growing usage of text messages. Text-based communication plays a critical role in various contexts, including connecting distant friends and family \cite{textBenefitInPandemic}, facilitating patient-doctor communication \cite{texthealthcare}, and enabling chatbot-supported businesses. However, text chats are inherently limited in conveying nonverbal affective information, leading to a decreased sense of connectedness and presence and an increased likelihood of miscommunication of emotional states \cite{stayConnected,MiscommunicationEmotion,emoBalloon}.

To compensate for this limitation, users often employ emojis or emoticons alongside their messages. HCI research has also explored more alternative methods of communicating emotions in textual messages, such as modifying typefaces or incorporating vibrations and bio-signals \cite{emotype,animatedtexts,vibemoji,heartRateMessageLiu}. A few recent studies have investigated using the color or shapes of chat balloons for conveying emotions \cite{emoBalloon,bubbleColoring}. While chat balloons appear to be a promising affective medium for textual messages, little is known about whether or how we could design animations of chat balloons to convey the more concrete and dynamic aspects of emotional expressions. And this has served as the primary motivation for our design-driven exploration.

In this paper, we present AniBalloons, a set of chat balloon animations designed to communicate six types of emotions: Joy, Anger, Sadness, Surprise, Fear, and Calmness. Following a structured affective animation design process \cite{kineticharts} and design requirements specifically tailored for chat balloons, we analyzed 230 affective animation examples to extract design patterns for each emotion category and iteratively designed five animations for each category. We then conducted a study assessing the affect recognizability of the designed animations and their perceived emotional properties according to the valence-arousal emotion model. Our research objective is to understand to what extent the intended emotion could be discerned from the designs and collect the nuanced emotional parameters for better understanding the affective affordance of the design animations. 

Our study results indicate that 80\% of the designed animations effectively communicate the intended emotions without support from textual messages; and the animation designs cover a variety of valence-arousal parameters, suggesting a great potential for chat-balloon animations as a unique affective channel for text messages. These findings open up a wide range of opportunities for future design and research for a range of text message-based systems across various devices, including social messaging and chatbot interfaces.

\section{Background}
In the realm of affective enhancement for text messages, a variety of approaches have been explored to compensate for the inherent limitations in nonverbal emotional communication. The majority of research has focused on emoticons or emojis \cite{ScottFahlman,plato}, with numerous studies examining their usage, user-led creativity, and the continuous introduction of new designs and customization capabilities \cite{Zhang2017_affectivestates,Hagen2019_emojiuse,Wiseman2018_repurposingemoji,kelly2015characterising,newemoji2021,facebookmessenger,Telegram}. In addition to emoticons, researchers have investigated the use of images, such as static or animated memes \cite{messageImage,jiang2017_GIF,Griggio2021}, and the visual features of texts, including typefaces and text animations \cite{emotype,emotionalSubtitles}, to convey emotions more effectively. However, the universal "container" of messages, chat balloons, has received comparatively less attention, presenting ample opportunities for further design and research.

Chat balloons or text bubbles, ubiquitous in text-based communication interfaces, have their roots in comics and manga where they served not only as carriers for speech but also as emotional indicators: e.g., using various shapes of text balloons \cite{balloonShape}. Recently, EmoBalloon by Aoki et al. \cite{emoBalloon} explored using explosion-shaped balloons to convey emotional arousal. Alternatively, Chen et al. \cite{bubbleColoring} investigated using colors to represent different sender affects. While these two studies highlight the potential of chat balloons as a means to complement nonverbal affective cues, the use of chat balloon animations for conveying emotions remains largely under-explored. Our study tackles this opportunity.

The use of animation to convey emotions has a long history in animation theory \cite{thomas1995illusion,louToLife,lasseter1987principles} and has been employed in various fields such as user interface design \cite{chevalier2016animations} and data visualization \cite{de2017taxonomy}. Animations, with their additional temporal dimension, allow for richer and more nuanced affective states that cannot be easily conveyed through static media. HCI research has explored the design of animations for GUI components, such as Kineticons \cite{Harrison2011}, which provided a set of animation designs for a wide range of GUI elements. Similarly, Kineticharts \cite{kineticharts} designed a collection of animation effects to enhance the affective expressiveness of data stories. Our research, inspired by these developments, aims to provide a collection of affective animations for chat balloons, and formally assess these animations for their affect recognizability and emotional properties, in order to inspire their usage across various message-based communications.

\section{design of AniBalloons}
The design of AniBalloons lasted for over a year. Based on the theories of basic emotions by Ekman \cite{ekman1992argument,ekman1992there}, we decided to start our design from six commonly communicated types of emotions: joy, anger, fear, surprise, sadness, and calmness (in the future more nuanced types of emotions could be added). A structured design workflow inspired by Kineticharts by Lan et al. \cite{kineticharts} was used. As the first step, we collected design inspirations from popular platforms like Dribbble, Behance, and Pinterest, resulting in a total of 336 motion graphic designs. After refining the inspirations and excluding designs that relied mainly on texts or static facial expressions (which were deemed less useful in terms of extracting animation design patterns), 230 examples remained. An open coding method was adopted by two designers in the research team to analyze the inspirations and extract affective animation design patterns based on three aspects: the main object's motion, decorative dynamic effects, and timing. Three designers from the research team then iteratively designed chat balloon animations for the six emotion categories using the extracted design patterns. The team focused on translating animation patterns into motions of a generic rounded rectangular chat balloon (so that the designed animations could be applied to a wide range of text bubbles across message-based applications), balancing expressiveness and unobtrusiveness. Expert review sessions with three professional animation designers (one from Canada and two from Asia) were conducted to further develop and refine the designs. The final design collection included five representative animations for each emotion category, polished and optimized based on expert feedback.

As the current outcome of the Aniballoons project, we developed 30 unique animations representing six different emotion categories (see \textbf{Appendix A} and the \textbf{supplementary video}): \textbf{Anger:} These designs utilize tension, body squeezing, and dynamic effects like fire-blowing, volcano eruptions, and explosions to convey anger. \textbf{Calmness:} They often involve metaphors related to water, air, and a floating or weightless state to evoke a sense of calm. \textbf{Fear:} They mainly use trembling motions and dynamic effects like fluctuating silhouettes and shaking dashed lines to express fear. \textbf{Joy:} They employ motions like jumping, stretching, and swinging, often accompanied by effects such as confetti and shining stars. \textbf{Sadness:} They convey sadness through crying-related motions and effects like tears, sobbing, and collapsing or melting. \textbf{Surprise:} These designs utilize sudden appearance or proximity through effects like zooming, splashing, and exclamation marks.

\section{Evaluation of Aniballoons}
In our evaluation study of Aniballoons, we aimed to address whether or how the designed chat balloon animations could convey emotions and how people would perceive their emotional properties. To achieve this, we conducted a study with 40 participants to evaluate the 30 designed animations with two objectives: \textbf{Affect Recognizability} (if participants could identify the intended emotion conveyed by each chat balloon animation without any hints from the message content) and \textbf{Emotional properties} (how the designs are perceived in terms of valence and arousal. This information helps identify the range of emotions that Aniballoons can effectively convey to complement other emotional cues such as emoticons).

\textbf{Stimuli:} The 30 Aniballoons designs were used as stimuli, rendered on grey chat balloons to eliminate the influence of base color. Dynamic decorative effects retained their original colors, as they would in actual usage. To avoid message content influence, pseudo-Latin texts (Lorem ipsum) were used as the text placeholder inside the chat balloons (see \textbf{Appendix A} for a number of examples).

\textbf{Procedure:} A total of 40 participants aged 18-44 (95\% aged 18-34) were recruited, with a gender distribution of 42.5\% women and 57.5\% men. A web-based survey interface was employed for the evaluation, where participants provided consent and demographic information before viewing the 30 designs in a randomized order. For each design, participants were asked to identify which of the six types of emotions the sender was conveying (an option of ``Other'' with a user input field is also given), which is an evaluation technique also used in \cite{kineticharts} for affect recognizability test. Meanwhile, participants were also asked to rate their perception of valence (pleasant-unpleasant) and arousal (calm-exciting) using two seven-point scales. Eight participants were randomly selected for a 15-minute follow-up remote interview to discuss their survey responses.

\section{Findings and Discussion}
\subsection{Affect Recognizability}
\begin{figure*}[!tb]
  \centering
  \includegraphics[width=\linewidth]{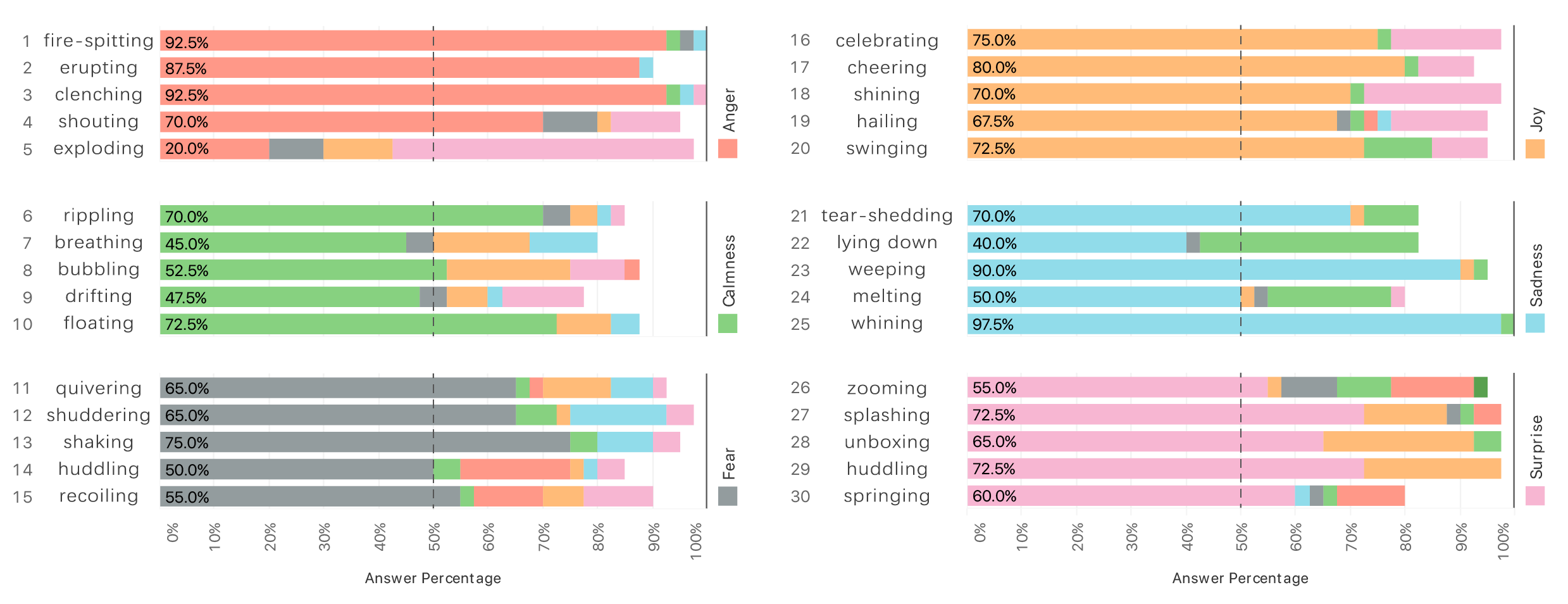}
  \vspace{-8mm}
  \caption{Evaluation results of AniBalloons' affect recognizability}

  \label{fig:recognizability}
\end{figure*}

The evaluation of Aniballoons' affect recognizability yielded positive results. As \autoref{fig:recognizability} shows, 80\% (24 out of 30) of the animations achieved a recognition accuracy greater than 50\% (without any hint from the message), a threshold considered effective in affective design as per prior works \cite{kineticharts,Ma2012GuidelinesFD}. Notably, 20 designs reached an accuracy of 65\% or higher. All the animations from the categories of Joy and Surprise were successfully recognized by the majority of participants, with four out of five animations from Fear and Anger categories, and three out of five from Sadness and Calmness categories, achieving similar success.

However, six animations did not exceed the 50\% accuracy mark. These included \textit{\#5-Exploding} from Anger (20\%), \textit{\#7-Breathing} (45\%) and \textit{\#9-Drifting} (47.5\%) from Calmness, \textit{\#14-Huddling} (50\%) from Fear, and \textit{\#22-Lying-down} (40\%), and \textit{\#24-Melting} (50\%) from Sadness. Despite this, upon considering participants' inputs with the ``Other'' option, and counting their interpretations that were closely related to the intended emotion, the success rate increased to 90\% (27 out of 30 animations). For instance, \textit{Drifting} was interpreted as ``meditating'', ``emotionless'', and ``sleepiness'', and \textit{Melting} was interpreted as ``disappointment'' or ``upset'', which were all close to their respectively intended emotions (Calmness and Sadness).

\subsection{Emotional Properties}
\begin{figure*}[!t]
  \centering
  \includegraphics[width=0.45\linewidth]{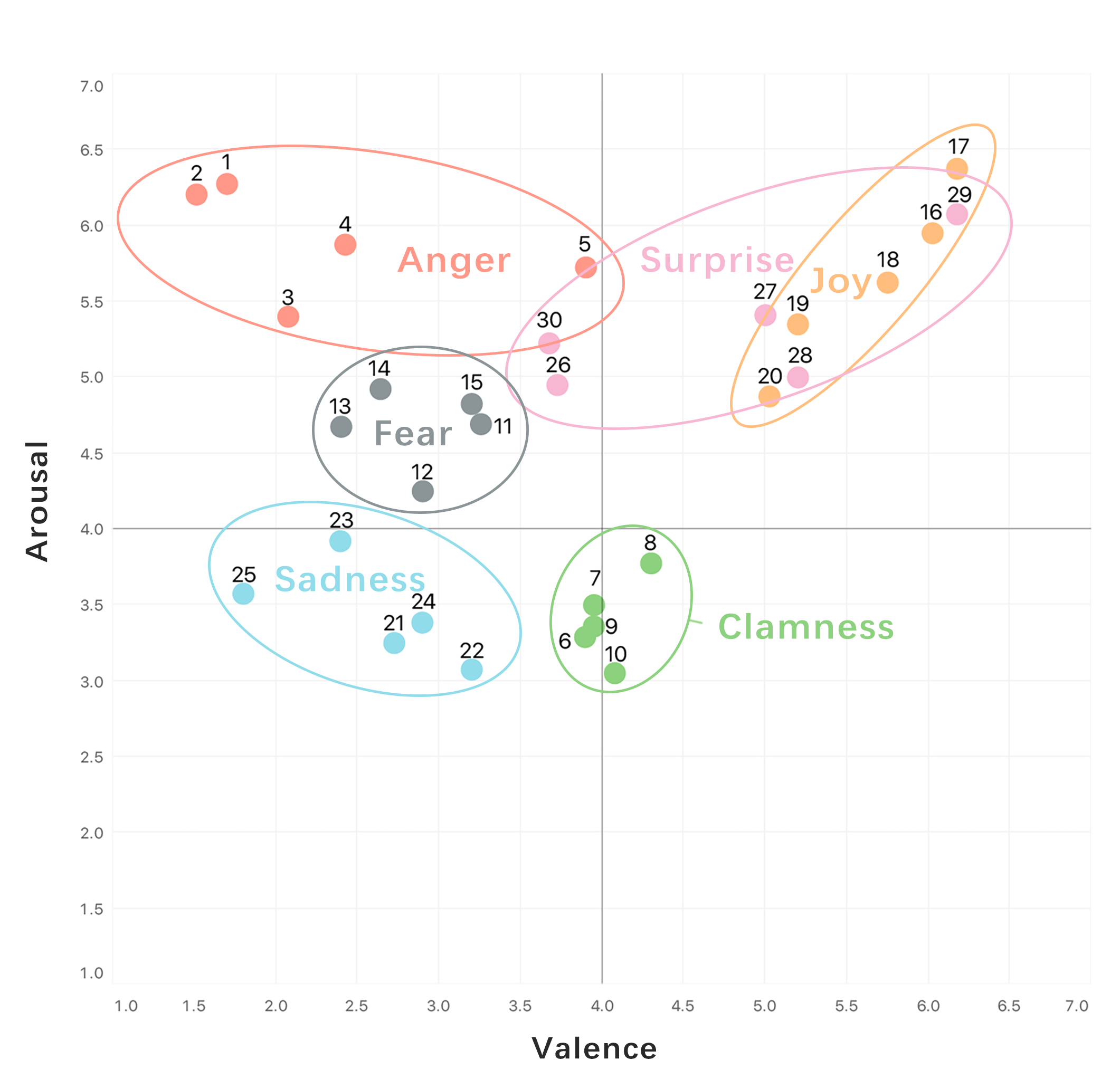}
  \includegraphics[width=0.45\linewidth]{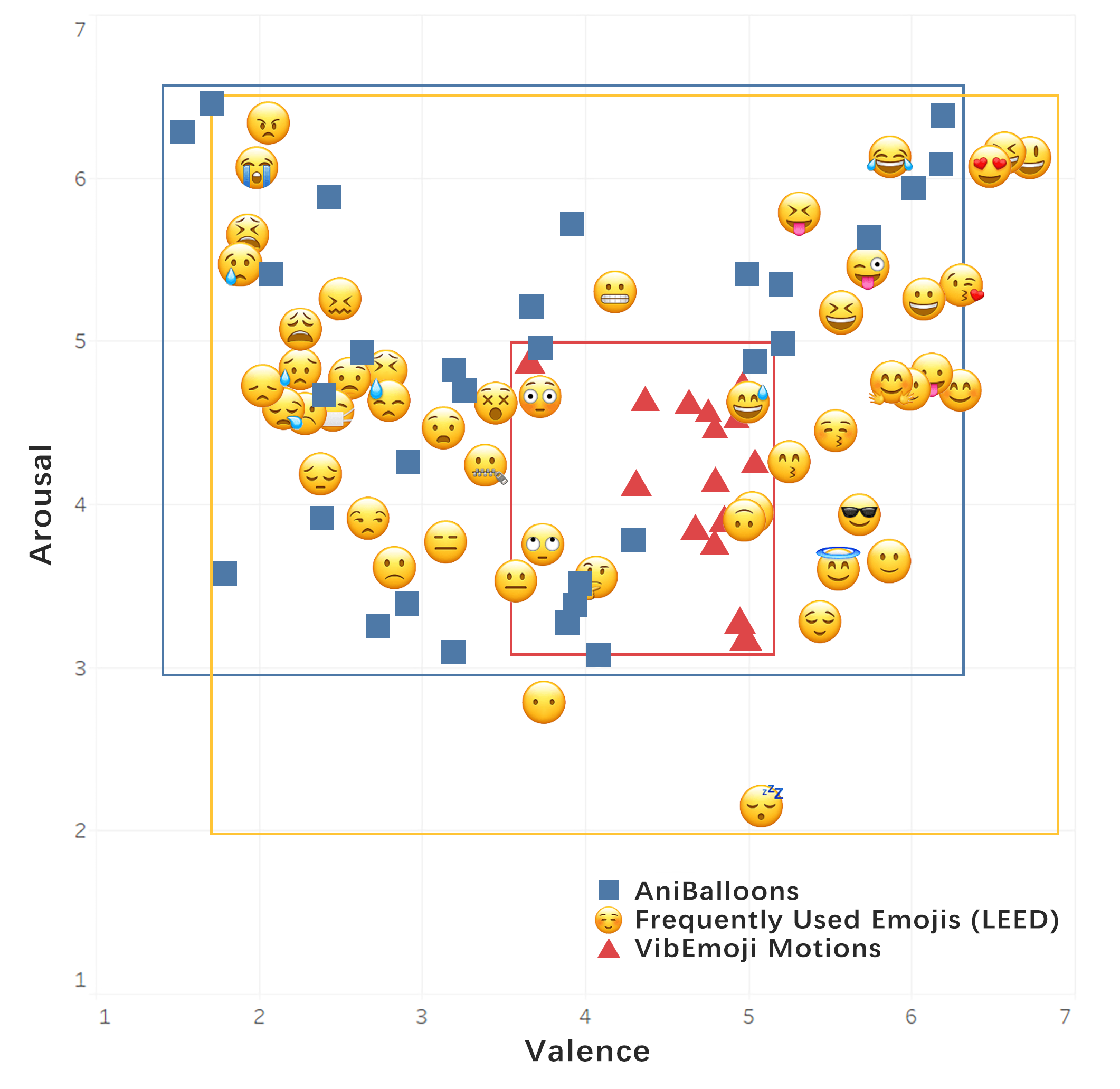}
  \vspace{-3mm}
  \caption{Evaluation results of AniBalloons' emotional properties. LEFT: AniBalloons' valence-arousal distribution. RIGHT: AniBalloons' valence-arousal distribution in reference with frequently used emojis, as well as motions effects of VibEmoji \cite{vibemoji}.}

  \label{fig:Properties}
\end{figure*}

Based on the valence-arousal data gathered, we mapped out all 30 designs of AniBalloons on a valence-arousal plane. As illustrated in \autoref{fig:Properties}, the visualized results revealed that each emotion category occupied a specific range on the plane, affording rich options for expressing nuanced variations of a specific emotion.

As shown in \autoref{fig:Properties}-LEFT, Joy-based designs spread throughout the positive-arousing section, with \textit{\#17-Cheering} being the most positive-arousing, while \textit{\#20-Swinging} was the least. The Anger and Fear designs resided in the negative-arousing section, with \textit{\#2-Erupting} and \textit{\#13-Shaking} expressing the strongest negative feelings. The Sadness designs lay in the negative-calm section, with \textit{\#25-Whining} perceived as the most negative. The Surprise designs spanned neutral and positive regions, with \textit{\#28-Unboxing} conveying the highest positivity, while \textit{\#26-Zooming} and \textit{\#30-Springing} reflected valence-neutral surprise. Calmness designs were located around the neutral region on the valence axis, reflecting their goal of expressing a feeling free from strong emotions.

Comparing Aniballoons' distribution with frequently used emojis and a prior set of animations used for multimodal emoticons showed that Aniballoons cover a wide range of emotional parameters, comparable to frequently used emojis and broader than the VibEmoji animations \cite{vibemoji} (which similarly focuses on affective communication and gathered same emotional property data). This suggests that Aniballoons can support diverse affective expressions in daily communication. However, there is potential for more designs in the positive-calm phase to enhance affective expression in this area.

Interview data from participants emphasized how Aniballoons can complement existing methods of emotional expression, such as emojis and stickers. Participants noted that Aniballoons' animations are straightforward, less likely to be misinterpreted, and can seamlessly integrate into any message with a chat balloon. Interestingly, the participants also mentioned a unique benefit of animated chat balloons as relatively more abstract expressions of emotions than emojis. Namely, unlike emojis or stickers, which are often selected to match personal characteristics such as skin color, gender, or culture, Aniballoons were seen as more universal due to their abstract designs based on motion patterns.

\section{Concluding Remarks and Future Opportunities}
This paper has introduced AniBalloons, a novel form of chat balloon animations designed to facilitate affective communication in text-based conversations. The design of AniBalloons was guided by a comprehensive process that involved extracting design patterns from affective animation examples and mapping them to six commonly conveyed emotions: Joy, Surprise, Sadness, Fear, Anger, and Calmness.

An evaluation study involving 40 participants was conducted to assess the effectiveness of AniBalloons in conveying intended emotions (without a hint from the message content) and to understand their perceived emotional properties. The results showed that 80\% of the designed animations successfully conveyed the intended emotions, with 90\% achieving this success when considering related interpretations. These results highlight the potential of AniBalloons to enhance the emotional expressiveness of message-based communications.

The study also revealed the nuanced emotional properties of the designed animations and their distribution on the valence-arousal plane. The distribution of AniBalloons animations was found to cover a wide range of emotional parameters, on par with frequently used emojis. This suggests the potential of AniBalloons to support a broad spectrum of affective expression in everyday communication contexts.

Moreover, the study confirmed the initial design intents of AniBalloons. Participants recognized the dynamic nature of AniBalloons as a complement to static emojis and appreciated the seamless integration of AniBalloons into any message that has a chat balloon. Furthermore, the abstract nature of AniBalloons' designs was appreciated for their universality, which makes them not restricted to specific demographics or personal characteristics.

Looking ahead, the insights gained from this research offer potential directions for future work. There is an opportunity to expand the repertoire of AniBalloons to cover more nuanced emotions and to explore the design space in the positive-calm phase of the emotional space. Furthermore, since the participants considered AniBalloons to be a complement to emojis and stickers, future research could explore the potential of combining these modalities to create richer and more nuanced emotional expressions. Lastly, as AniBalloons were designed based on extracted motion patterns and abstract effects that were heavily based on shape formation, the methodology could be extended to other animation components, such as color changing or motion speed, opening up new avenues for enhancing emotional expression in digital communication.




\begin{acks}
We thank all the participants and collaborators in this study and all reviewers of this paper. This work is supported in part by the Waterloo-Huawei Joint Innovation Lab and The NSSFC Art Grant (22CG184).
\end{acks}

\bibliographystyle{ACM-Reference-Format}
\bibliography{references}

\appendix
\section{AniBalloons}
\autoref{fig:AniBalloons} presents the 30 designs of AniBalloons in static frames (see the supplementary video for the dynamic version). \autoref{fig:greyballoons} exemplifies the animated chat balloons used for evaluation.
\begin{figure*}[b]
  \centering
  \includegraphics[width=0.95\linewidth]{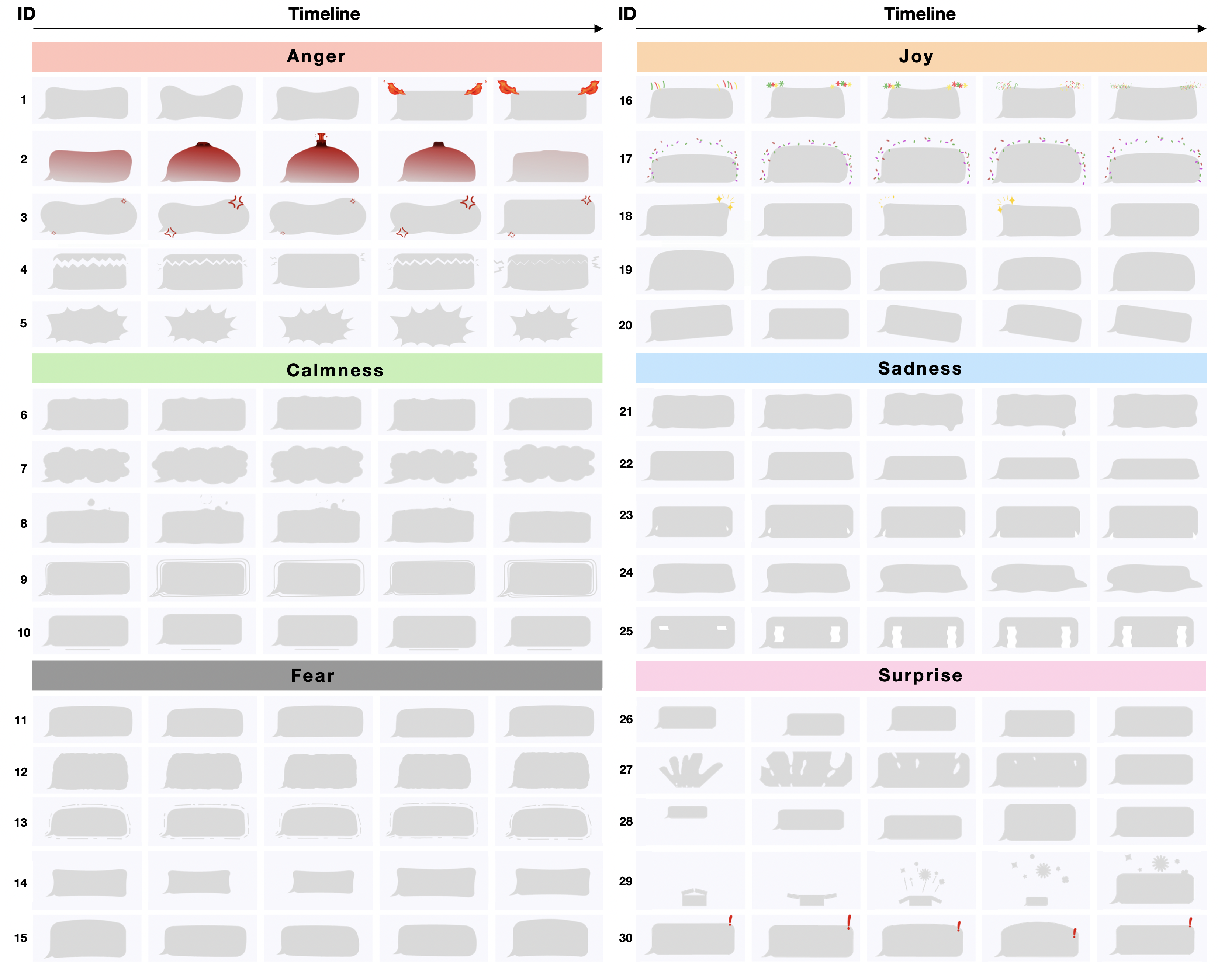}
  \vspace{-4mm}
  \caption[caption]{
  The 30 designs:\\
  {\small{\tabular[t]{rlllll}
  \textbf{Anger:} & 1-Fire-spitting & 2-Erupting & 3-Clenching & 4-Shouting & 5-Exploding\\
  \textbf{Calmness:} & 6-Rippling & 7-Breathing & 8-Bubbling & 9-Drifting & 10-Floating\\
  \textbf{Fear:} & 11-Quivering & 12-Shuddering & 13-Shaking & 14-Huddling & 15-Recoiling\\
  \textbf{Joy:} & 16-Celebrating & 17-Cheering & 18-Shining & 19-Hailing & 20-Swinging\\
  \textbf{Sadness:} & 21-Tear-shedding & 22-Lying-down& 23-Weeping& 24-Melting &25-Whining\\
  \textbf{Surprise:} & 26-Zooming & 27-Splashing & 28-Popping & 29-Unboxing & 30-Springing
  \endtabular}}}
  \label{fig:AniBalloons}
\end{figure*}

\begin{figure*}
  \centering
  \includegraphics[width=\linewidth]{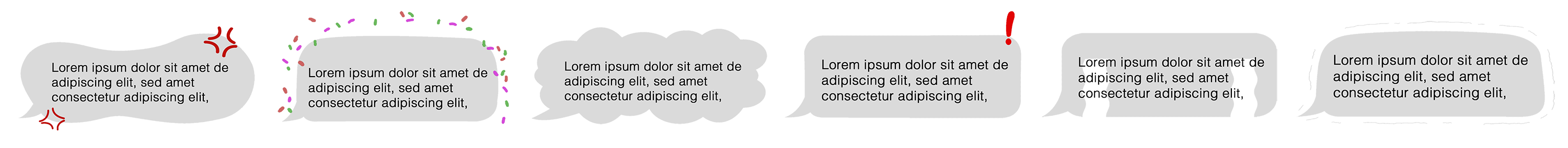}
  \vspace{-8mm}
  \caption{examples of animated chat balloons used for evaluation: rendered on chat balloons with grey base color and pseudo-Latin texts}
  \label{fig:greyballoons}
\end{figure*}

\end{document}